\documentclass[prl,floatfix,twocolumn,showpacs,amsmath,amssymb]{revtex4}
\usepackage{graphicx}
\usepackage{dcolumn}
\usepackage{bm}
\newcommand{\dagga}{{\phantom{\dagger}}}
\begin{document}
\title{Variational description of Mott insulators} 

\author{Manuela Capello,$^{1,2}$ Federico Becca,$^{1,2}$ Michele Fabrizio,$^{1,2,3}$
Sandro Sorella,$^{1,2}$ and Erio Tosatti$^{1,2,3}$}
\affiliation{
$^{1}$  International School for Advanced Studies (SISSA), Via Beirut 2-4, 
I-34014 Trieste, Italy \\
$^{2}$ INFM-Democritos National Simulation Centre, Trieste, Italy. \\
$^{3}$ International Centre for Theoretical Physics (ICTP), P.O. Box 586, I-34014 Trieste, Italy
}

\date{\today}

\begin{abstract}
The Gutzwiller wave function for a strongly correlated model can, if supplemented
with a long-range Jastrow factor, provide a proper variational description of 
Mott insulators, so far unavailable. We demonstrate this concept in the prototypical 
one-dimensional $t-t^\prime$ Hubbard model, where at half filling we reproduce
all known phases, namely the ordinary Mott undimerized insulator with power-law 
spin correlations at small $t^\prime/t$,  the spin-gapped metal above a 
critical $t^\prime/t$ and small $U$, and the dimerized Mott insulator at 
large repulsion. 
\end{abstract}


\maketitle

Since Mott's original proposal~\cite{mott} the correlation-driven metal-Mott 
insulator transition (MIT) has attracted rising interest, renewed by the 
discovery of novel strongly correlated materials. On the verge of becoming 
Mott insulators, many systems display very unusual properties, high-T$_c$ 
superconductivity being one spectacular example. While understanding Mott 
insulators and MITs is conceptually simple, calculations constitute a hard 
and long standing problem.
Conventional electronic structure methods, such as Hartree-Fock (HF) 
or density functional theory in the local density approximation (LDA) cannot
describe MITs, unless one allows for some kind of symmetry breaking. 
The standard example is long-range static magnetic order in the unrestricted 
HF, local-spin-density, or LDA+U approximations. This device works by 
effectively turning the MIT into a 
conventional metal-band insulator transition, thus masking the essence 
of the Mott phenomenon, where a charge gap appears quite independently of 
spin order. The fact that most known Mott insulators are indeed accompanied 
at low temperatures by some symmetry breaking, usually of magnetic type,
further encourages the (wrong) surmise that it is not possible to describe any 
Mott insulator without a symmetry breaking. 

Another useful and popular approximation that may invite the same conclusion 
is based on the variational Gutzwiller wave function (GWF) and its various 
generalizations.~\cite{gutzwiller,shiba,baeriswyl,gebhard} 
The GWF is the simplest way to improve a symmetry-unbroken, hence metallic,  
Slater determinant by partly projecting out the expensive double-occupancy
charge configurations. In principle, were the projection complete, 
the GWF would indeed describe a Mott insulator devoid of symmetry breaking. 
Full projection however means zero band-energy gain, generally incorrect, 
except for infinite on-site repulsion. For finite projection, appropriate
at finite repulsion, the GWF unfortunately describes a metallic state in any 
finite dimensional lattice, at least so long as the uncorrelated Slater 
determinant state is metallic.~\cite{shiba} 
To obtain an insulator, one is forced once again to Gutzwiller project an 
artificially symmetry-broken determinant wave function (WF).
The main drawback of the GWF can be immediately recognized if one recalls   
Mott's original description of a correlation-driven insulator. Let us consider 
for simplicity the single-band Hubbard model at half-filling. First of all 
it is clear that the system must even in the insulating phase allow for 
charge fluctuations around the mean value, implying some doubly-occupied 
(doublon, D) and empty (holon, H) sites, or it would jeopardize all 
band-energy gain. 
However, in order to describe an insulator, D and H have to be bound,
otherwise any infinitesimal electric 
field could induce electric current. This spatial correlation among expensive 
charge configurations is exactly what is missing in the conventional GWF. 
There have been several attempts to cure this defects 
by allowing H-D correlations just on nearest neighbor 
sites,~\cite{shiba2,penc} with only partial success. It is actually 
suggested by our previous discussion that long-range H-D correlations 
should be a fundamental ingredient of a realistic insulating WF, otherwise 
there would always exist a finite probability for H and D to escape 
from one another, signaling of a metallic behavior.~\cite{ogata}

In this Letter we show that this plan can be accomplished with success. 
In particular, we demonstrate for a specific one-dimensional (1D) model 
that: {\sl (i)} non-trivial insulating WFs with or without 
symmetry breaking do exist; {\sl (ii)} the MIT, so far well described 
only in infinite dimensions,~\cite{georges} is accessible variationally 
even in finite dimensions.  
We consider the 1D $t-t^\prime$ Hubbard model 
\begin{eqnarray}
{\cal H} &=& -t \sum_{i,\sigma} c^\dag_{i,\sigma} c^\dagga_{i+1,\sigma} +H.c.
+t^\prime \sum_{i,\sigma} c^\dag_{i,\sigma} c^\dagga_{i+2,\sigma} +H.c. \nonumber \\
&+&U \sum_i n_{i,\uparrow} n_{i,\downarrow},\label{Ham}
\end{eqnarray}
where $c^\dag_{i,\sigma}$ ($c^\dagga_{i,\sigma}$) creates (destroys) an electron
with spin $\sigma$ on site $i$ and
$n_{i,\sigma} = c^\dag_{i,\sigma} c^\dagga_{i,\sigma}$. In the following
we will assume $t$ and $t^\prime$ positive.

\begin{figure}
\includegraphics[width=\columnwidth]{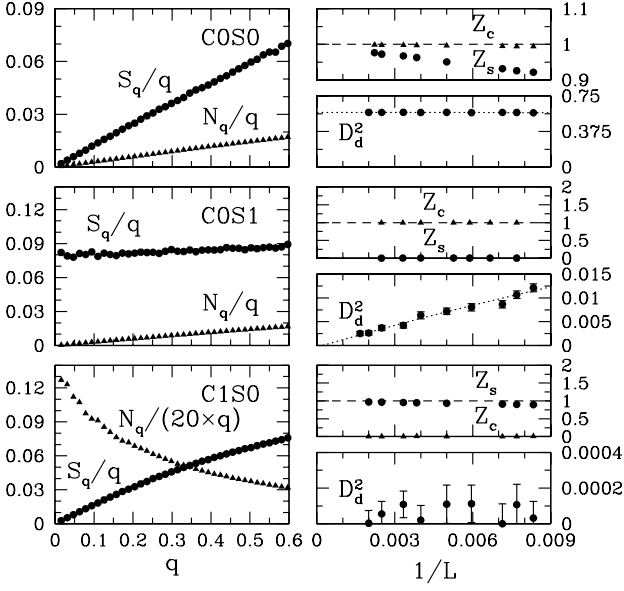}
\vspace{-0.5cm}
\caption{\label{fig:wf}
Spin and charge properties of the $C0S0$ WF (top panels),
the $C0S1$ WF (middle panels), and the $C1S0$ WF (bottom panels).
Left panels: $N_q/q$ (triangles) and $S_q/q$ (circles) as a function 
of $q$ for $L=400$.
Right panels: $D_d^2$, $Z_c$, and $Z_s$ as a function of $1/L$. 
The dotted line is a three parameter fit and the dashed line marks the value 1.
The parameters of the WF are: $Am=0.33$, $A^2=0.17$ ($C0S0$ and $C0S1$), 
$m=0$, $A^2=8.4\times 10^{-3}$ ($C1S0$), $\Delta_1=2t$, $\Delta_2=1.7t$
($C1S0$ and $C0S0$), and $\Delta_1=\Delta_2=\Delta_3=0$ ($C0S1$).
}
\end{figure}

Bosonisation~\cite{fabrizio} and density-matrix 
renormalization group calculations~\cite{arita,noack} predict that 
the ground state at half-filling is an insulator with gapless spin 
excitations (labeled $C0S1$, where $CnSm$ indicates a state with $n$ 
gapless charge modes and $m$ gapless spin modes) for $t^\prime/t \alt 0.5$, 
a spin-gapped metal ($C1S0$) with strong superconducting fluctuations 
for $t^\prime/t \agt 0.5$ and small $U/t$, and a 
fully gapped spontaneously dimerized insulator ($C0S0$) for 
$t^\prime/t \agt 0.5$ and large $U/t$. 
Dimerization is characterized by
the discrete order parameter
$D_d^2= \lim_{|i-j| \to\infty}
\vert \chi(i-j-1)-2\chi(i-j)+\chi(i-j+1) \vert$,
where $\chi(i-j) = 9 \langle S_i^z \cdot S_{i+1}^z\;
S_j^z\cdot S_{j+1}^z \rangle$ gives   the dimer-dimer correlation
function of  spin rotationally invariant WFs,
$S_i^z$ being the spin operator along z-axis  at site $i$.

\begin{figure}
\includegraphics[width=0.4\textwidth]{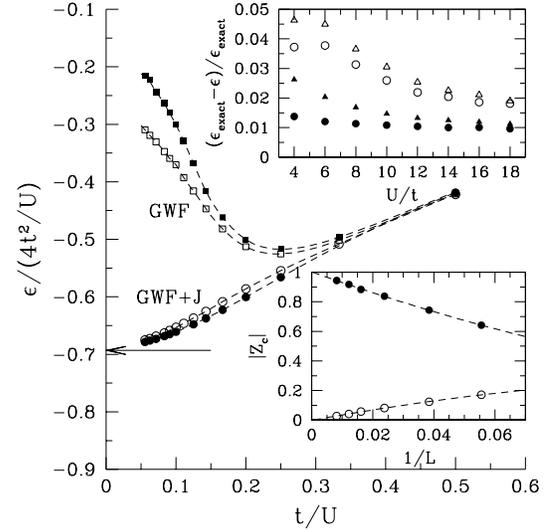}
\vspace{-0.2cm}
\caption{\label{fig:strongcoupling}
Energy per site $\epsilon$ for the simple 
GWF ($L=18$, empty squares, and $L=82$, full squares) and 
for the WF with long-range density-density Jastrow (GWF+J)
($L=18$, empty circles, and $L=82$, full circles), for $t^\prime=0$. 
The Slater determinant is a simple Fermi sea, with $\Delta_q=0$. 
The arrow indicates the exact energy per site of the Heisenberg model and
the lines are guides to the eye.
Top inset: accuracy of the WF with all the density-density Jastrow
independently minimized (same symbols as before) and with the analytic
parametrization ($L=18$, empty triangles, and $L=82$, full triangles).
Bottom inset: $Z_c$ for the GWF+J (full circles) 
and for the GWF (open circles), lines are three parameter fits.
}
\end{figure}

The $C1S0$ metallic phase suggests a variational WF built out of BCS, namely 
\begin{equation}
|\Psi \rangle = {\cal P} \, |BCS \rangle = 
{\cal P} \, {\rm exp} 
\left( \sum_{q} f_{q} \, c^{\dag}_{q,\uparrow} c^{\dag}_{-q,\downarrow} \right)
|0\rangle,\label{WF}
\end{equation}
where $|BCS \rangle$ is the ground state of a BCS Hamiltonian  
with gap function $\Delta_q$ and dispersion
$\epsilon_q=-2 \cos(q) +2t^\prime \cos(2q)-\mu$, $\mu$ being the free electron 
chemical potential. The pairing function is defined by 
$f_q= \Delta_q/(\epsilon_q+E_q)$, with the BCS energy spectrum
$E_q=\sqrt{\epsilon_q^2+\Delta_q^2}$, and
$\Delta_q=\Delta_1 \cos(q) +\Delta_2 \cos(2q) +\Delta_3 \cos(3q)$, 
$\Delta_1$, $\Delta_2$, and $\Delta_3$ being variational parameters.
Obviously $|BCS\rangle$ reduces to the Fermi sea for $\Delta_q=0$.
The operator ${\cal P} ={\cal J} {\cal P}_N$, where
${\cal P}_N$ projects onto the subspace with fixed number of electrons 
$N=L$ and ${\cal J}$ is the Jastrow factor:
\begin{equation}
{\cal J}={\rm exp} \left(  \sum_{i,j} \frac{1}{2}\,v_{i,j}\, (n_i-1)(n_j-1) 
+ w_{i,j}\, h_i d_j \right),\label{Jastrow}
\end{equation}
where $n_i=n_{i,\uparrow}+n_{i,\downarrow}$ is the density operator, 
$d_i=n_{i,\uparrow}n_{i,\downarrow}$ the doublon operator and 
$h_i=(1-n_{i,\uparrow})(1-n_{i,\downarrow})$ the holon one. 
We remark that, in presence of a magnetic flux $\phi\not=0$, the Jastrow term 
can acquire a non-centro-symmetric component, $w_{i,j}-w_{j,i}\not = 0$. 
While the centro-symmetric part $w_{i,j}+w_{j,i}$ turns out 
to be less crucial than $v_{i,j}$ 
for the $\phi=0$ ground state, the odd component 
$w_{i,j}-w_{j,i}$, which is zero at zero flux and had not been included
so far,~\cite{Millis} will play a relevant role in 
determining the response to $\phi$.
In practice we find it more convenient to discriminate the Mott 
insulator from the metal through correlation functions which can 
be calculated in absence of flux, as it is very unhandy
to follow the changes of the variational WF due the presence of $\phi$.

Numerically we find that the optimized $v_{i,j}$ is smooth as a function
of the distance $|i-j|$, while $w_{i,j}$ is staggered, 
$w_{i,j}=|w_{i,j}|\, (-1)^{|i-j-1|}$. 
Although $w_{i,j}$ improves the accuracy, it does not change qualitatively 
the nature of the WF, which is mainly determined by $v_{i,j}$, as we discuss below. 
For this reason most of the results are obtained with $w_{i,j}=0$.
Indeed, since $(n_i-1)(n_j-1) = h_i h_j +d_i d_j -h_i d_j -d_i h_j$, then  
$v_{i,j}<0$ implies H-H (D-D) long-range repulsion   
and H-D long-range attraction. The latter embodies 
the binding of H and D, while the repulsion prevents accumulation of H-D pairs. 
That is just the desired type of correlations missing in the GWF. 

We notice that optimization of the  
Jastrow factor allows to fulfill variationally the f-sum rule. Indeed, if we 
denote the variational energies of $|\Psi \rangle$ by ${\cal E}_0$ and
of $n_q |\Psi \rangle$ by ${\cal E}_q$, then we can prove   
for the optimized $v_{i,j}$'s that 
\begin{equation}   
\lim_{q\to 0} \, ({\cal E}_q-{\cal E}_0) = \frac{q^2}{2\,N_q}\,  
\frac{\langle\Psi|\sum_{k,\sigma} \partial^2_k \epsilon_k\, 
c^\dag_{k,\sigma} c^\dagga_{k,\sigma} |\Psi\rangle}{\langle\Psi|\Psi\rangle},
\label{f-sum-rule}
\end{equation}
where $N_q=\langle \Psi|n_{-q} n_q|\Psi \rangle /\langle \Psi|\Psi \rangle$
is the equal-time density-density correlation function. On the assumption 
that our variational WF $|\Psi\rangle$ gives a fair description of 
the actual ground state WF, 
Eq.~(\ref{f-sum-rule}) allows to infer the behavior of the long-wavelength 
charge excitation spectrum by just analysing the charge structure factor $N_q$ 
of $|\Psi\rangle$. Namely, if $N_q \sim q$, there should exist gapless 
charge modes, while if $N_q \sim q^2$ charge excitations are presumably gapped.
In turn $N_q$ is determined by the structure factor of the uncorrelated 
BCS WF, $N^0_q$, and by the Fourier transform of $v_{i,j}$, $v_q$, 
through $N_q\sim N^0_q/(1+N^0_q v_q)$.~\cite{reatto} Since $N^0_q$ tends 
to a constant value for $q\to 0$, the long wavelength behavior of $N_q$ 
is actually determined by the singular behavior of $v_q$, $v_q \sim 1/q$ or 
$v_q \sim 1/q^2$ implying $N_q\sim q$ or $N_q\sim q^2$, respectively. 
In other words the long-range Jastrow factor not only restores gauge 
invariance, $\lim_{q\to 0}\,N_q=0$, but also filters out either the 
metallic or the insulating component of the uncorrelated WF by an unbiased
minimization of the energy.
 
Despite the large number of variational parameters, 
$v_{i,j}$'s, $w_{i,j}$'s as well as 
$\Delta$'s, it is possible to optimize the full WF by  
recent developments in the variational Monte Carlo 
technique.~\cite{sorella} 
Yet, to afford larger scale simulations,  
we also assume an analytic parametrization of $v_{i,j}$:
\begin{equation}
v_{i,j} = - \frac{1}{L}\sum_{q} \, \cos\left[q(i-j)\right]\,  
\frac{A}{\sqrt{\sin^2(q/2)+m^2}-m},\label{v_q-ansatz}
\end{equation}
with $A$ and $m$ variational parameters, with $m$ playing the role of the 
inverse Mott-localization length: $m=0$ or $m>0$ implying respectively metallic, 
$v_q\sim 1/q$, or insulating, $v_q\sim 1/q^2$, behavior.

Besides the small-$q$ behavior of the charge structure factor $N_q$, 
another quantity which can discriminate between a metal and an insulator 
is a ``rigidity'' defined through 
$Z_c = \langle \Psi| 
{\rm exp} \left( \frac{2\pi i}{L} \sum_j j \, n_j \right) 
|\Psi \rangle / \langle \Psi|\Psi \rangle$.
In the thermodynamic limit, for 1D systems, $|Z_c| \to 0$ for a metal 
and $|Z_c| \to 1$ for an insulator.~\cite{resta}

For what concerns the spin modes, we argue that their behavior is determined 
by the BCS spectrum $E_q$, namely if $E_q$ is gapless then 
also the spin excitations are gapless and the spin structure factor 
$S_q\sim q$, whereas, if ${\rm min}(E_q)>0$, the spin excitation spectrum 
has a gap and $S_q\sim q^2$. This is supported by the direct 
calculation of the generalized spin ``rigidity''
$Z_s = \langle \Psi|
{\rm exp} \left( \frac{4\pi i}{L} \sum_j j \, S_j^z \right)
|\Psi \rangle / \langle \Psi|\Psi \rangle$, $|Z_s| \to 0$ for a spin-gapless
WF and $|Z_s| \to 1$ for a spin-gapped WF.

First of all we now prove that the WF~(\ref{WF}) with appropriate 
choice of parameters is able to describe  
$C0S1$, a Mott insulator with no symmetry breaking, as well as $C0S0$, 
a spontaneously dimerized insulator, and $C1S0$, a spin-gapped metal.  
In Fig.~\ref{fig:wf}, we show how the three possible cases $C0S1$, $C1S0$, 
and $C0S0$ can be actually realized.
The two WFs corresponding to $C0S0$ and $C0S1$ are indeed insulating even
though the average density of H (D) is finite.  
Moreover, we note that, unlike the uncorrelated BCS WF, 
the correlated variational WF in the $C1S0$ phase is not superconducting,
as it should of course be in 1D, thanks to the long-range Jastrow 
factor.~\cite{dagotto}
Remarkably, even if the $C0S0$ BCS WF is not explicitly dimerized, after 
application of the Jastrow factor~(\ref{Jastrow}), a finite dimerization 
order parameter emerges. 
As in the liquid-solid transition of helium, it is typical of Jastrow functions
to describe phase transitions in two-body correlation functions, without
an explicit symmetry breaking in the one-body properties.
 
\begin{figure}
\includegraphics[width=0.4\textwidth]{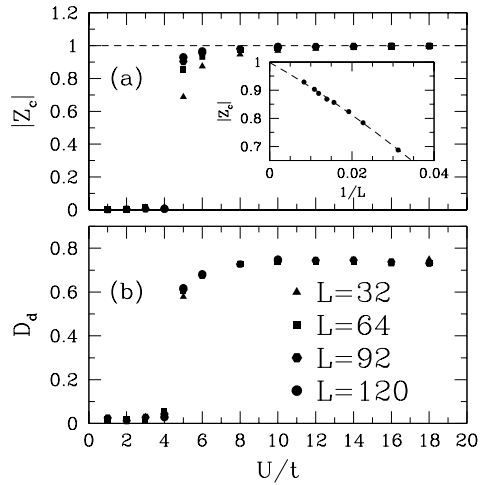}
\vspace{-0.5cm}
\caption{\label{fig:resta}
(a) $|Z_c|$ as a function of $U/t$ for $t^\prime/t=0.75$ and different sizes.
(b) The same for the dimer order parameter $D_d$.
The $v_{i,j}$'s are independently minimized, very similar results are
obtained with the parametrization of Eq.~(\ref{v_q-ansatz}).
In the inset: the size scaling of $|Z_c|$ for $U/t=5$, the dashed line is a 
three parameter fit.
}
\end{figure}

To check the quality of the variational WF~(\ref{WF}) for the 
Hamiltonian~(\ref{Ham}), we start by considering 
the case of $t^\prime=0$, which allows us a 
direct comparison with exact Bethe {\it ansatz} results. 
For the reasons we mentioned previously, the GWF gives a rather poor 
variational description of the half-filled Hubbard model for $U \gg t$. 
In this limit an alternative procedure is commonly adopted. It is known that 
for $U\gg t$ the half-filled Hubbard model transforms by a Schrieffer-Wolff 
transformation ${\rm e}^{i\, {\cal{S}}}$ into an Heisenberg model. 
Therefore, instead of using the GWF, it is more convenient 
to use ${\rm e}^{i\, {\cal{S}}}\, |\Psi_G\rangle$, where $|\Psi_G\rangle$ 
is the {\em fully projected} GWF. In that case, one searches for the 
optimal uncorrelated WF which, after complete projection, minimizes the 
average of ${\rm e}^{-i\, {\cal{S}}}\,{\cal{H}}\, {\rm e}^{i\, {\cal{S}}}$, 
namely of an Heisenberg Hamiltonian. Practically 
one is forced to expand 
${\rm e}^{i\, {\cal{S}}}\simeq 1 + i\,{\cal{S}}$, which limits
the applicability of this WF to the large $U/t$ regime.~\cite{gros}
On the contrary our WF is equally 
accurate both at weak and at large $U/t$, as shown in Fig.~\ref{fig:strongcoupling}. Remarkably the long range Jastrow correlations 
turn a fully metallic Fermi sea- the worst case one electron band for an insulator- into a perfectly good Mott insulator. 

\begin{figure}
\vspace{-0.6cm}
\includegraphics[width=0.4\textwidth]{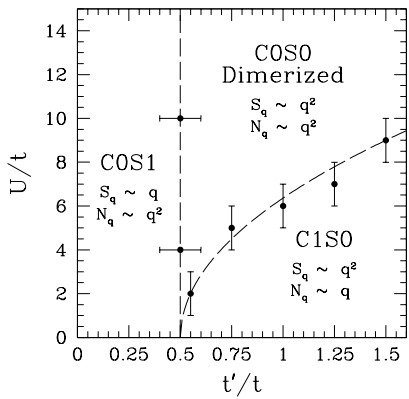}
\vspace{-1.25cm}
\caption{\label{fig:phase}
Variational phase diagram of the 1D $t{-}t^\prime$ Hubbard model. The
error bars keep into account finite-size effects and dashed lines are guides
to the eye.
}
\end{figure}

Going back to the $t-t^\prime$ Hubbard model, in Fig.~\ref{fig:resta} 
we show the behavior of $|Z_c|$ and 
$D_d$ for $t^\prime/t=0.75$.
By increasing $U/t$, $|Z_c|$ shows a sudden increase, and we can easily 
discriminate between the metallic phase, where $|Z_c| \to 0$ by increasing $L$,
and an insulating phase, where $|Z_c| \to 1$ for $L \to \infty$. Moreover,
in the latter case, the MIT is indeed accompanied by a spontaneous 
dimerization, marked by a finite dimer order parameter $D_d$, see 
Fig.~\ref{fig:resta}(b).

The variational phase diagram for the half-filled $t{-}t^\prime$ Hubbard
model is finally shown in Fig.~\ref{fig:phase}.
In the region of $t^\prime/t \alt 0.5$ and $U>0$, we find no evidence 
of a phase transition, apart from finite-size effects at small $U/t$. 
The best variational state has $N_q \sim q^2$, 
indicating a charge gap and $S_q \sim q$, gapless spin excitations:
$\Delta_q$ connects only different sublattices (i.e., $\Delta_2=0$), 
making $E_q$ gapless.
For $t^\prime/t \agt 0.5$, there is a clear MIT at finite $U/t$ between a spin-gapped 
metal, stable for small $U/t$ and a dimerized insulator, stable at large $U/t$.
In the metallic phase, the variational WF has $N_q \sim q$, 
$E_q$ gapped, with 
$\Delta_2 \ne 0$, although $|\Delta_q|$ is small, hence 
$S_q \sim q^2$, corresponding to an exponentially decaying spin-spin
correlations.
By increasing $U/t$ and entering into the insulating phase there is a fast 
increase of $\Delta_2$,  with $E_q$ always fully gapped, and 
$v_q \sim 1/q^2$. In this phase therefore $N_q \sim q^2$ and $S_q \sim q^2$.
Although the variational WF does not explicitly break translational
symmetry, there is a finite dimerization in the
thermodynamic limit. Remarkably, this 
is due to the concomitant effect of a singular Jastrow
$v_q \sim 1/q^2$ and a gapped BCS spectrum $E_q$.

In conclusion, it is possible to describe a Mott 
insulator by a variational WF based on a metallic Fermi sea  
which   includes, besides a  partial Gutzwiller 
projector, a long-range density-density Jastrow term. The
accuracy of the WF is verified in a one-dimensional $t-t^\prime$ Hubbard 
model at half-filling, which has a non trivial phase diagram comprising
different phases and a MIT. 
We argue that similar variational WFs with $v_q \sim 1/q^2$ could describe 
non-trivial Mott insulating phases in higher dimensions too, giving
the possibility to improve realistic descriptions of Mott insulating 
compounds.  We also believe these 
wave functions to hold a great potential promise for future attacks to 
the superconducting phases that result by doping Mott insulators. 

This work was partially supported by INFM, by MIUR (COFIN 2003), by FIRB 
RBAU017S8R004, and FIRB RBAU10LX5H. F.B. was supported by an INFM research 
position. We acknowledge useful discussions with C. Castellani, A. Parola,
and G. Santoro.


\begin{thebibliography}{99}

\bibitem{mott} N.F. Mott, Proc. Phys. Soc. (London) {\bf 62}, 416 (1949).
\bibitem{gutzwiller} M.C. Gutzwiller, \prl {\bf 10}, 159 (1963).
\bibitem{shiba} H. Yokoyama and H. Shiba, J. Phys. Soc. Jpn. {\bf 56}, 
  1490 (1987).
\bibitem{baeriswyl} M. Dzierzawa {\it et al.}, Helv. Phys. Acta {\bf 70},
  124 (1997).
\bibitem{gebhard} F. Gebhard, \prb {\bf 41}, 9452 (1990).
\bibitem{shiba2} H. Yokoyama and H. Shiba, J. Phys. Soc. Jpn. {\bf 59}, 
  3669 (1990).
\bibitem{penc} P. Fazekas and K. Penc, Int. J. of Mod. Phys. B {\bf 1}, 
  1021 (1988).
\bibitem{ogata} See however: H. Yokoyama {\it et al.}, cond-mat/0308264.
\bibitem{georges} A. Georges {\it et. al}, \rmp {\bf 68}, 13 (1996).
  G. Kotliar {\it et al.} \prl {\bf 87} 186401 (2001).  
  Within dynamical mean-field theory, or its cluster generalizations, 
  the counterpart of real-space holon-doublon binding is a binding in the 
  time domain, a simple manifestation of the Mott-Hubbard gap.
\bibitem{fabrizio} M. Fabrizio, \prb {\bf 54}, 10054 (1996).
\bibitem{arita} R. Arita {\it et al.}, \prb {\bf 57}, 10324 (1998);
  K. Kuroki {\it et al.}, J. Phys. Soc. Jpn. {\bf 66}, 3371 (1997).
\bibitem{noack} S. Daul and R.M. Noack, \prb {\bf 61}, 1646 (2000).
\bibitem{Millis} A.J. Millis and S.N. Coppersmith, \prb {\bf 43}, 13770 (1991).
\bibitem{reatto} L. Reatto and G.V. Chester, Phys. Rev. {\bf 155}, 88 (1967).
\bibitem{sorella} S. Sorella, \prb {\bf 64}, 024512 (2001).
\bibitem{resta} R. Resta, \prl {\bf 80}, 1800 (1998).
\bibitem{dagotto} S. Sorella {\it et al.}, \prl {\bf 88}, 117002 (2002).
\bibitem{gros} C. Gros {\it et al.}, \prb {\bf 36}, 381 (1987).

\end{thebibliography}
\end{document}